\documentclass[american]{article}
\usepackage[T1]{fontenc}
\usepackage{amsmath}
\usepackage{amssymb}
\usepackage{graphicx}
\usepackage{esint}
\usepackage{authblk}
\usepackage{color}
\usepackage{babel}
\usepackage{enumerate}
\usepackage{cite}
\usepackage{lineno}

\usepackage{tikz}
\usetikzlibrary{shapes.geometric, arrows}

\makeatletter

\providecommand{\tabularnewline}{\\}

\@ifundefined{definecolor}
 {\usepackage{color}}{}
\@ifundefined{definecolor}{\usepackage{color}}{}

\usepackage{babel}

\makeatother
\makeatletter
\renewcommand*{\@fnsymbol}[1]{\ensuremath{\ifcase#1\or \dagger\or *\or \ddagger\or
   \mathsection\or \mathparagraph\or \|\or **\or \dagger\dagger
   \or \ddagger\ddagger \else\@ctrerr\fi}}
\makeatother

\usepackage{babel}
\begin{document}
\title{Alpha Cluster Model of Atomic Nuclei}

\author[1]{Zbigniew Sosin\thanks{deceased}}
\author[2]{Jan B\l{}ocki}
\author[1]{Jinesh Kallunkathariyil\thanks{jinesh.kallunkathariyil@gmail.com}}
\author[3]{Jerzy~\L{}ukasik}
\author[3]{Piotr Paw\l{}owski}
\affil[1]{M. Smoluchowski Institute of Physics, Jagiellonian University, \L{}ojasiewicza 11, 30-348 Krak\'{o}w, Poland}
\affil[2]{NCBJ, Theoretical Physics Division (BP2), \'{S}wierk, Poland}
\affil[3]{IFJ PAN, Krak\'{o}w, Poland}
\maketitle 
\begin{abstract}

Description of a nuclear system in its ground state and at low excitations based
on the equation of state (EoS) around normal density is presented. In the
expansion of the EoS around the saturation point additional spin polarization
terms are taken into account. These terms, together with the standard symmetry
term, are responsible for appearance of the $\alpha$-like clusters in the ground
state configurations of the N=Z even-even nuclei. At the nuclear surface these
clusters can be identified as alpha particles. A correction for the surface
effects is introduced for atomic nuclei. Taking into account an additional
interaction between clusters the binding energies and sizes of the considered
nuclei are very accurately described. The limits of the EoS parameters are
established from the properties of the $\alpha$, $^{3}$He and $t$ particles. 

\end{abstract}

\section{Introduction}

In 1936 Bethe and Bacher \cite{Bethe1936} and in 1938 Hafstad and Teller
\cite{Hafs1938} predicted that alpha particle structures \cite{Web,Freer2010} 
could be present in atomic nuclei. Convincing arguments for existence of such
structures were provided by systematics of the binding energies of the even-even
nuclei with equal number of protons and neutrons (see also  
\cite{vonoertzen06}), as well as by systematics of the binding energy of the
additional neutron in nuclei like $^{9}$B, $^{13}$C and $^{17}$O. The latter
systematics could be explained by assuming that the valence neutron moves in a
multi-center potential with centers identified with alpha particles. Today, in a
commonly accepted point of view one considers that alpha particle clusters can
appear in the nuclear matter at subsaturation densities. Such conditions are met
e.g. at the nuclear surface where, in not too heavy nuclei, almost the whole
nuclear mass is concentrated.

Subtle changes of the ground state energy of atomic nuclei as a function of
nucleon number can be described by the shell model, which neglects correlations
between nucleons. The main idea of the present approach is to verify whether
these subtle changes, in case of $n$-alpha nuclei, could be attributed to the
appearance of alpha-like clusters which modify the structure of the wave
function and introduce strong correlations between nucleons. In such an approach
the system in its ground state behaves like a crystal, with stationary
configuration and shape and with definite distances between the wave function
centers for a given nucleus. The subtle changes of the binding energy due to
appearance of alpha-like clusters in the wave functions of $n$-alpha nuclei will
be determined within an extended version of the model proposed in
\cite{Sos_2014}. 

It will be shown in the following that the hypothesis of alpha structures in the
$n$-alpha nuclei can indeed describe the binding energy systematics of Bethe
\cite{Bethe1936,Hafs1938}. It will be also shown that the hypothesis of the
alpha structures helps to understand the evolution of the root mean square (rms)
radii and of the density profiles as a function of the number of alpha particles
in the $n$-alpha nuclei.

The paper is organized as follows: section 2 describes the adopted interactions
in the nuclear matter. Section 3 presents constraints on the introduced
parametrization resulting from the properties of light charged particles.
Section 4 is devoted to the ground state properties of $n$-alpha nuclei. First, it
investigates the effects of clusterization on the ground state configurations.
Then, some corrections to the Hamiltonian, related to the many body character of
the nuclear interaction and to the bosonic nature of the wave function are
introduced. It will be shown that the introduced corrections make the structural
corrections stronger. Finally, the model predictions for binding energies, rms
radii and density profiles are presented and compared with the experimental
values. Section 5 contains conclusions.

\section{Interactions in nuclear matter around saturation}

In order to predict correctly a state and evolution of a nuclear system
a reliable description of the interaction between nucleons is needed. In case of a
strong interaction it is not an easy task. As we know the fundamental
theory governing the nucleon-nucleon interactions is the quantum chromodynamics
(QCD), where interactions are calculated from the physics of quarks
and gluons. Unfortunately, very limited progress has been achieved
in application of QCD for the description of interaction between quark-gluon
clusters. Therefore a phenomenological approach, associated
to some specific conditions has been applied. Since we are interested
in the low-energy region in which nucleons are not excited internally,
one can treat nucleons as quasi elementary particles. Additionally,
if we apply a non-relativistic approximation, the mutual nucleon interaction
can be described by a respective potential. As can be found in the text
books, details of the form of such a potential are deduced from nuclear
interaction symmetries and mechanism of exchange of bosons.

For a final test of this type of a potential and the determination of
its parameters one uses a comparison with the experimental data. Here
the nucleon-nucleon scattering data or light nuclei properties are
usually used. A general conclusion from this kind of research is that a
simple two body potential is not sufficient to describe the behavior
of the system of more than two nucleons. One should add at least the
three-body force. In particular, in the nuclear system, the three-body
force is acceptable because the strong interaction is generated by
the exchange process of gluons. The n-body interaction can be interpreted
as coming from the multi-pion exchange processes.

These considerations show the difficulties in the description of even
the simplest nuclear system. The extension of the description
of interaction by the three-body forces shows that the exchanged particles
(bosons) play a significant role in the interaction process.

In order to describe dynamics of the system one has to express the change of
energy associated with its evolution. In the description of energy through the
potential one has to define its form. Hence the wave function of a system of
identical particles must be either symmetric or antisymmetric and it must be
taken into account when defining the energy of the ground state. In case of the
fermionic systems one should use antisymmetric wave function which is a serious
difficulty in the description of the collision dynamics. Some proposals to solve
this problem are shown for instance in \cite{Feld_90,Ono_92,Kanada12}. Other approaches,
introducing some kinds of the momentum dependent Pauli potentials and
Pauli-blocking schemes, which are meant to mimic the effects of
antisymmetrization and are much less time consuming, are also proposed, see e.g.
\cite{wile1,dors1,hori1,boal1,maru1}.

Therefore, in the present approach, we propose a description in which particles
are distinguishable and energy associated with the fermionic motion affects the
resulting energy density which is parametrized and used instead of the
potential. Such a concept is present in the description of the liquid drop model
(LDM) and in some density functional models, e.g. in \cite{fink}. It is obvious
that such an approach requires the knowledge of the local density of matter and
the knowledge of the wave function describing the system. In the present
description we use a standard one, commonly applied in such cases, i.e. the wave
function is defined as a simple product of minimal wave packets (eq. (1)), as
e.g. in \cite{Aich91}. The main argument for disregarding the antisymmetrization
in the present approach is the form of the proposed EoS which includes the spin
and isospin terms. These terms force the unlike nucleons to form ``bosonic''
alpha-like clusters for which, presumably, the antisymmetrization may no longer
play a decisive role. 

Each minimal wave packet in the adopted wave function has a defined spin and a
specific charge for a selected quantization axis: 

\begin{equation}
\Phi={\displaystyle \prod_{k=1}^{A}}\,^{k}\phi_{I_{k}S_{k}}\label{eq:wf}
\end{equation}

\begin{equation}
\,^{k}\phi_{I_{k}S_{k}}=\frac{1}{\left(2\pi\sigma_{k}^{2}(r)\right)^{3/4}}\exp\left(\frac{-\left(\mathbf{r}_{k}-\left\langle
\mathbf{r}_{k}\right\rangle
\right)^{2}}{4\sigma_{k}^{2}(r)}+\frac{i}{\hbar}\mathbf{r_{k}}\left\langle
\mathbf{p}_{k}\right\rangle \right)\label{eq:wf_k} \end{equation}
 where
$\sigma_{k}^{2}(r)$, $\left\langle \mathbf{r}_{k}\right\rangle $, $\left\langle
\mathbf{p}_{k}\right\rangle $, are the width (the position variance of the
$k$-th nucleon) of a Gaussian wave packet and the mean position and mean
momentum of each of the $A$ nucleons, respectively. Every partial wave function
is labeled by $I_{k}=n\;\mathrm{or}\; I_{k}=p$ and
$S_{k}=\uparrow\;\mathrm{or}\; S_{k}=\downarrow$, denoting the isospin and spin
(or precisely, their projections on a specified quantization axis) associated
with a given nucleon. Variables $\left\langle \mathbf{r}_{k}\right\rangle
$, $\left\langle \mathbf{p}_{k}\right\rangle $ and $\sigma_{k}^{2}(r)$ are, in
general, time-dependent parameters describing the wave functions.

In the present approach the nuclear matter is treated as a four component
fluid characterized by the respective densities:

$\rho_{p\uparrow}$  - for protons with spin up,

$\rho_{p\downarrow}$ - for protons with spin down,

$\rho_{n\uparrow}$   - for neutrons with spin up,

$\rho_{n\downarrow}$ - for neutrons with spin down. 

Density distributions in the system are uniquely determined by parameters of the
wave function (mean position, mean momentum and variance of the position of each
nucleon). The description of the system evolution due to the time evolution of
these parameters is, in general, defined by the Dirac-Frenkel time-dependent
variational principle \cite{Dirac81}. This approach describes the interaction in
a self-consistent way.

Assuming that nucleons are moving inside the nuclear system, the average energy
of the system is a sum of the average potential energy and of kinetic energy
associated with the fermionic motion. This is the case when one disregards the
energy of the ordered motion, which for the low excitations of the system is
considered later on as a correction (see formula (\ref{eq:Ham})).

According to the above idea we now present a method of determining the
average energy associated with each nucleon. For the description
of this average energy we assume the existence of a scalar field $\varepsilon\left(\rho_{p\uparrow},\rho_{p\downarrow},\rho_{n\uparrow},\rho_{n\downarrow}\right)$,
determined by the local densities $\rho_{p\uparrow}\!\!\left(\mathbf{r}\right),\rho_{p\downarrow}\!\!\left(\mathbf{r}\right),\rho_{n\uparrow}\!\!\left(\mathbf{r}\right),\rho_{n\downarrow}\!\!\left(\mathbf{r}\right)$.

If, for every $k-th$ nucleon, the corresponding wave packet
$^{k}\phi_{I_{k}S_{k}}$ determines the probability
$P_{k}(\mathbf{r})=\left|^{k}\phi_{I_{k}S_{k}}\right|^{2}$ of finding a nucleon
at a given point \textbf{$\mathbf{r}$} then it is assumed that the average
energy $e_{k}$ associated with this nucleon is defined by the mean value and
variance of the field
$\varepsilon\left(\rho_{p\uparrow},\rho_{p\downarrow},\rho_{n\uparrow},\rho_{n\downarrow}\right)$
and can be expressed as:

\begin{equation}
e_{k}=\left\langle \varepsilon\right\rangle _{k}+\lambda\sigma_{k}\left(\varepsilon\right)\label{eq: ave}
\end{equation}
 where: 
\begin{equation}
\left\langle \varepsilon\right\rangle _{k}=\int P_{k}(\mathbf{r})\varepsilon\left(\rho_{p\uparrow},\rho_{p\downarrow},\rho_{n\uparrow},\rho_{n\downarrow}\right)d^{3}\mathbf{r}\label{eq:ave_1}
\end{equation}
 
\begin{equation}
\sigma_{k}^{2}\left(\varepsilon\right)=\int P_{k}(\mathbf{r})\left(\varepsilon\left(\rho_{p\uparrow},\rho_{p\downarrow},\rho_{n\uparrow},\rho_{n\downarrow}\right)-\left\langle \varepsilon\right\rangle _{k}\right)^{2}d^{3}\mathbf{r}\label{eq:ave_2}
\end{equation}
 and $\lambda$ is a parameter related to the surface energy, 
similarly as in \cite{Sos_2010}.

The EoS of the nuclear matter is defined for an infinite system, assuming its
homogeneity and isotropy. For such a matter, variance of the associated field
$\varepsilon\left(\rho_{p\uparrow},\rho_{p\downarrow},\rho_{n\uparrow},\rho_{n\downarrow}\right)$
vanishes and the average value of the energy per nucleon is determined by the
mean value of the field alone. The field
$\varepsilon\left(\rho_{p\uparrow},\rho_{p\downarrow},\rho_{n\uparrow},\rho_{n\downarrow}\right)$
is in this case equivalent to the EoS of the nuclear matter. Thus, the
functional form of the field
$\varepsilon\left(\rho_{p\uparrow},\rho_{p\downarrow},\rho_{n\uparrow},\rho_{n\downarrow}\right)$
can be obtained from the EoS by applying the local density approximation.

In the present work we use a 12-parameter cubic form of the EoS proposed in
\cite{Sos_2014}. In the vicinity of the saturation density it can be
approximated using a second order Taylor expansion:

\[
e=e_{00}+\frac{K_{0}}{18}\xi^{2}+
\]
 
\[
\delta^{2}\left(e_{I0}+\frac{L_{I}}{3}\xi+\frac{K_{I}}{18}\xi^{2}\right)+
\]

\[
\left(\eta_{n}^{2}+\eta_{p}^{2}\right)\left(e_{ii0}+\frac{L_{ii}}{3}\xi+\frac{K_{ii}}{18}\xi^{2}\right)+
\]
 
\begin{equation}
2\eta_{n}\eta_{p}\left(e_{ij0}+\frac{L_{ij}}{3}\xi+\frac{K_{ij}}{18}\xi^{2}\right)\label{eq:eos_4-1}
\end{equation}
 where:

\begin{equation}
\xi=\frac{\rho-\rho_{0}}{\rho_{0}}\label{eq:ksi}
\end{equation}

\begin{equation}
\delta=\frac{\rho_{n}-\rho_{p}}{\rho}\label{eq:del}
\end{equation}
 
\begin{equation}
\eta_{n}=\frac{\rho_{n\uparrow}-\rho_{n\downarrow}}{\rho}\label{eq:n_n}
\end{equation}
 
\begin{equation}
\eta{}_{p}=\frac{\rho_{p\uparrow}-\rho_{p\downarrow}}{\rho}\label{eq:n_p}
\end{equation}
 in which $\rho$ and $\rho_{0}$ are the total nuclear matter density and
the density of isospin and spin balanced matter at saturation, respectively.

The first two terms in (\ref{eq:eos_4-1}) are just a standard, 6-parameter form
of the EoS approximation commonly used in the present day experimental and
theoretical studies, see e.g. \cite{bao}. The first term describes the symmetric
matter in a balanced system (zero isospin and spin), with $K_{0}$ being the
compressibility of the symmetric nuclear matter. The second one is a standard
approximation of the symmetry energy: 

\begin{equation}
e_{I}=e_{I0}+\frac{L_{I}}{3}\xi+\frac{K_{I}}{18}\xi^{2}\label{eq:esym_I}
\end{equation}
 with $e_{I0}$ being the Wigner constant and the coefficients $L_{I}$ and
$K_{I}$ being the slope and curvature, respectively.

The last two terms of (\ref{eq:eos_4-1}) are the main novelty of the present
approach (see also \cite{Sos_2014}). In analogy to the isospin symmetry energy,
they describe the spin symmetry energies for neutrons and protons separately: 

\begin{equation}
e_{ii}=e_{ii0}+\frac{L_{ii}}{3}\xi+\frac{K_{ii}}{18}\xi^{2}\label{eq:esym_ii}
\end{equation}
 and the energy of the mutual, spin interaction of protons and neutrons:

\begin{equation}
e_{ij}=e_{ij0}+\frac{L_{ij}}{3}\xi+\frac{K_{ij}}{18}\xi^{2}\label{eq:esym ij}
\end{equation}

The terms with indices $ii$, i.e. the $e_{ii0}$, $L_{ii}$ and $K_{ii}$,
(constant, slope and curvature) describe the energy of protons or neutrons.
Indices $ij$ indicate that the symmetry energy refers to the mutual interaction
of protons and neutrons. Determination of the expansion parameters in
(\ref{eq:eos_4-1}) will be discussed in the next chapters.

Using the field
$\varepsilon\left(\rho_{p\uparrow},\rho_{p\downarrow},\rho_{n\uparrow},\rho_{n\downarrow}\right)$
one can express the average value of the Hamiltonian for a given system of A
nucleons as: 

\begin{equation}
\left\langle \Phi\left|H\right|\Phi\right\rangle =\sum_{k=1}^{k=A}\frac{\left\langle \mathbf{p}_{k}\right\rangle ^{2}}{2m_{N}}+\sum_{k=1}^{k=A}\left\langle \varepsilon\right\rangle _{k}+\lambda\sum_{k=1}^{k=A}\sigma_{k}(\varepsilon)+\left\langle \Phi\left|V_{C}\right|\Phi\right\rangle \label{eq:Ham}
\end{equation}
 where $m_{N}$ is the nucleon mass and the last term describes the Coulomb
energy. As later on we will be searching for the ground state configurations of
the even-even nuclei with $N=Z$, it can be assumed that  the average momenta of
the wave packets $\left\langle \mathbf{p}_{k}\right\rangle $ are equal to zero
and, thus, the first term in the expression (\ref{eq:Ham}) can be neglected.
Following a standard approach (see e.g. \cite{bao}) it is assumed that the
parametrization of the EoS includes implicitly the kinetic contribution coming
from the Fermi motion. As has been demonstrated in \cite{Sos_2010}, the adopted
cubic approximation of the EoS is flexible enough to reproduce typical
predictions of sophisticated models of the nuclear matter which explicitly take
into account the kinetic contribution resulting from the Fermi motion.

Since the local density can be written as: 
\begin{equation}
\rho\left(\mathbf{r}\right)=\sum_{k=1}^{k=A}P_{k}\left(\mathbf{r}\right),\label{eq:dens_tot}
\end{equation}
 the average value of the Hamiltonian can be expressed as: 
\begin{equation}
\left\langle \Phi\left|H\right|\Phi\right\rangle =\int\varepsilon\left(\rho_{p\uparrow},\rho_{p\downarrow},\rho_{n\uparrow},\rho_{n\downarrow}\right)\rho\left(\mathbf{r}\right)d^{3}\mathbf{r}+\lambda\sum_{k=1}^{k=A}\sigma_{k}(\varepsilon)+\left\langle \Phi\left|V_{C}\right|\Phi\right\rangle \label{eq:Ham_int}
\end{equation}
 As one can see the Hamiltonian is a sum of three components, which
can be interpreted as a volume, surface and Coulomb energies. Therefore
the present description can be interpreted as a microscopic realization
of the LDM.

\begin{figure}[ht]
  \centering 
 \includegraphics[scale=0.4]{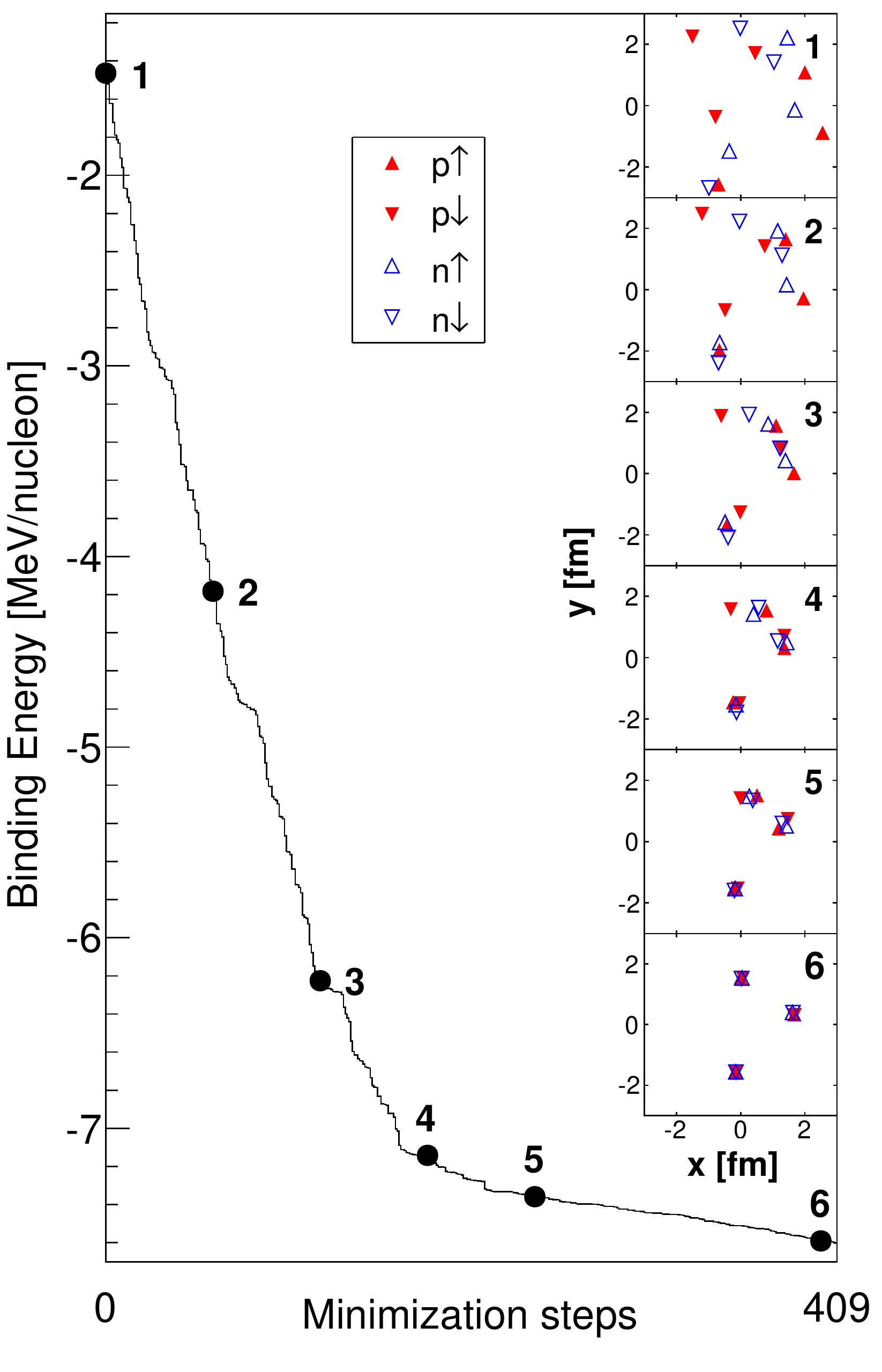}
 
\caption{Evolution of the binding energy of 12 nucleons (3$\times$p$\uparrow$,
3$\times$p$\downarrow$, 3$\times$n$\uparrow$ and 3$\times$n$\downarrow$) during
the minimization procedure. The insets 1-6 show the projections of the
configurations of nucleons on the x-y plane at selected stages of the
minimization. Panel 1 represents the  random initial configuration. Panel 6
shows the projection of the final ground state configuration of $^{12}$C.}

\label{figmin}
\end{figure}

Finding the ground state configuration of a group of A nucleons within the
adopted parametrization of the wave function and for given parameters of the
EoS, consists in finding the values of parameters $\sigma_{k}^{2}(r)$, 
$\left\langle \mathbf{r}_{k}\right\rangle $, and $\left\langle
\mathbf{p}_{k}\right\rangle$, which minimize the value of Hamiltonian.  The
procedure of finding the ground state configurations has been described in Sect.
4 of \cite{Sos_2010}. We would like to emphasize that the proposed
parametrization of the EoS, with two more terms in (6), is responsible for
driving  the nucleons of different types,  p$\uparrow$, p$\downarrow$,
n$\uparrow$, n$\downarrow$, to form quadruplets, i.e. to group into
$\alpha$-like clusters during the minimization procedure. These two terms play a
similar role as the standard symmetry term which attains a minimum when the
neutron-proton asymmetry $\delta$ of (7) tends to zero, i.e. when the
neutron-proton couples group together. This $\alpha$-like clustering property of
(6), is demonstrated in Figure \ref{figmin}.

The histogram in Fig. \ref{figmin} presents the evolution of the binding energy
of 12 nucleons during the process of searching for the ground state
configuration which minimizes the model Hamiltonian (16). The 6 panels of the
inset show ``snapshots'' of the projections of the mean positions of the nucleon
wave packets onto the x-y plane, at selected stages of the minimization
procedure. Panel 1 presents the initial random configuration, panels 2-5 some
intermediate ones and panel 6 shows the final ground state configuration of
$^{12}$C. As can be seen, the minimum of the Hamiltonian is achieved for a
configuration with three $\alpha$-like clusters. These clusters appear in the
corners of an equilateral triangle. Note, that in the figure the triangle is not
perfectly equilateral due to the 2-dimensional projection.

It has to be emphasized, that if clusters are allowed to appear in the
considered system the description has to be modified. Modifications of the
Hamiltonian due to clustering will be discussed in section 4.2.

\section{Constraints on the EoS from the properties of light charged particles}

Ground state properties of light charged particles, LCP, such as $d$, $t$, $^{3}$He or
$\alpha$ can provide important constraints on parameters which describe
interactions in the nuclear matter for at least four reasons. First, one can
assume that in their structure clusters of matter (obviously beyond nucleons)
can be formed. Therefore one may expect that clustering of matter has no effect
on their binding energies and sizes. Second, these LCPs do not contain more than
one $\alpha$ particle and thus one can neglect the $\alpha-\alpha$ interactions
(see Sect. 4.2) and use the simpler Liquid-Drop-like Hamiltonian (16). Third,
the measured density profiles of these light particles are relatively well
described by Gaussian distributions which is in line with the adopted
parametrization of the wave function. Fourth, simplicity of these particles
assures vanishing of some types of interactions in (6) and selective increase of
sensitivity to some other ones. For instance, parameters related to the isospin
symmetry energy (second term in (\ref{eq:eos_4-1})) in principle do not affect
the binding energy of a deuteron and alpha particle. Similarly, due to the
saturation of the proton (neutron) spin interactions in $^{3}$He ($t$), the
mutual spin interaction (fourth term in (\ref{eq:eos_4-1})) vanishes in their
case.

In case of LCPs we can assume that the average value of the
Hamiltonian is completely determined by the field
$\varepsilon\left(\rho_{p\uparrow},\rho_{p\downarrow},\rho_{n\uparrow},\rho_{n\downarrow}\right)$,
by the Coulomb interaction and by the surface energy. This assumption may not be
valid for more complex systems in which mutually interacting clusters can be
formed, eg. for $^{8}$Be$\rightarrow\alpha-\alpha$ or
$^{6}$Li$\rightarrow\alpha-d$. For such systems one can expect additional
interactions between clusters due to the specific many-body forces. The mean
value of the Hamiltonian (16) depends on 13 parameters: 12 parameters of the
field $\varepsilon(\mathbf{r})$ plus the surface tension parameter $\lambda$.
Binding energies and rms radii of the ground states for $d$, $t$, $^{3}$He and
$\alpha$ provide 8 equations, thus insufficient number of constraints for a
unique choice of parameters describing the interaction. In the present section
we will try to deduce a subspace of this 13-parameter-space for which the above
mentioned properties of LCPs are reproduced.

\subsection{Constraints on $\rho_{0}$, $e_{00}$, $K_{0}$
and $\lambda$ from the properties of an alpha particle}

It is clear that the greatest reduction of the number of parameters can be
expected in case of an alpha particle. Here, the spin and isospin dependent
interactions are practically absent. Thus, properties of alpha particles should
be described by four parameters only: $\rho_{0}$, $e_{00}$, $K_{0}$ and
$\lambda$. Constraints on these parameters are imposed by the rms radius and the
binding energy of an alpha. These parameters cannot be determined unambiguously.
However, there is a flexibility in determining a two-dimensional hypersurface,
in the space $\rho_{0}$, $e_{00}$, $K_{0}$ and $\lambda$, in which the searched
values of these parameters exist.

In order to determine this hypersurface the following $\chi^{2}$ variable is defined: 
\begin{equation}
\chi^{2}=\left(\frac{B_{\alpha e}-B_{\alpha m}}{B_{\alpha e}}\right)^{2}+\left(\frac{r_{\alpha e}-r_{\alpha m}}{r_{\alpha e}}\right)^{2}\label{eq:ch2}
\end{equation}
 where, $B_{\alpha e},\: B_{\alpha m},\: r_{\alpha e},\: r_{\alpha m}$ are the
experimental and model values of the binding energy and rms radius,
respectively. Note that a variable defined this way should be zero in the 
searched subspace.

\begin{figure}[h!]
    \centering
 \includegraphics[scale=0.35]{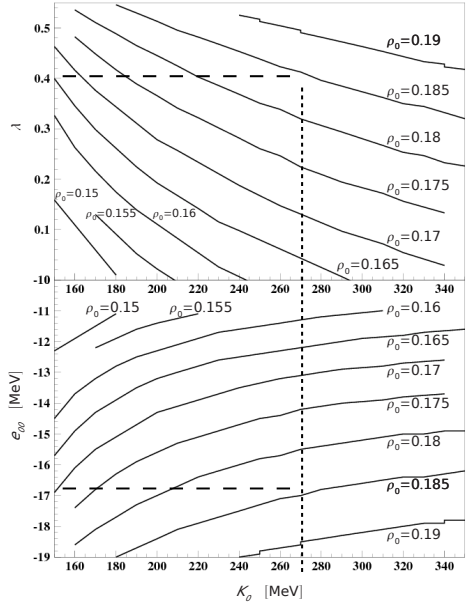}
 
\caption{Constant saturation density curves, $\rho_{0}$, as a function of
$e_{00}$, $K_{0}$ and $\lambda$, constrained by the ground state properties
of an $\alpha$ particle.}
\label{fig1}
\end{figure}

In order to determine the correlations among the parameters in a searched
subspace, the variables $\rho_{0}$ and $K_{0}$ are predefined on a grid, and the
remaining two parameters are found by requiring that $\chi^{2}<10^{-7}$ for the
model ground state configurations of an alpha particle. The results are shown in
Fig. \ref{fig1}.

The figure presents the constant saturation density curves as a function of the
other three parameters specified on the axis, which ensure a correct binding
energy and rms radius of an alpha particle for the model ground state
configurations. The dotted and dashed lines show how one can determine the
parameters $\lambda$ and $e_{00}$ for a given compressibility $K_{0}$ and
saturation density $\rho_{0}$. In the illustrated example the parameters
$\rho_{0}=0.185$ fm$^{-3}$ and $K_{0}=270$ MeV determine the values of
$\lambda=0.4$ and $e_{00}=-16.79$ MeV. These values of the parameters have been
suggested by additional studies of the competition between Coulomb, surface and
volume energies in the binding energy calculated for the ground state
configurations of the heavier nuclei (see Table 1). This problem will be
considered later on.

\subsection{Constraints on $e_{I0}$, $L_{I}$, $K_{I}$, $e_{ii0}$, $L_{ii}$ and
$K_{ii}$ from the properties of $t$ and $^{3}$He}

After fixing the parameters $\rho_{0}$, $e_{00}$, $K_{0}$ and $\lambda$ let us
try to examine the other ones. For $t$ and $^{3}$He the ground state energy does
not depend on the choice of parameters $e_{ij0}$, $L_{ij}$ and $K_{ij}$. This
results from vanishing of the $\eta_{n}\eta_{p}$ product in the last term of
(\ref{eq:eos_4-1}). Thus, the binding energies and rms radii for $t$ and
$^{3}$He can provide four equations for six unknown EoS parameters: $e_{I0}$,
$L_{I}$, $K_{I}$, $e_{ii0}$, $L_{ii}$ and $K_{ii}$, and so, two equations are
still missing. The two free parameters can be constrained from the properties of
heavier nuclei as it was done in the case of the fully symmetric nuclear matter.
At this stage it is only possible to examine the subspace defining the limits
for the considered parameters from the properties of the $t$ and $^{3}$He
nuclei. We notice the fact that the parameters associated with isospin:
$e_{I0}$, $L_{I}$ and $K_{I}$ differentiate between the binding energy and the
rms radii of $t$ and $^{3}$He while the parameters $e_{ii0}$, $L_{ii}$ and
$K_{ii}$ affect equally the energy and the size of these particles.

\begin{figure}[h]
  \centering 
  \includegraphics[scale=0.3]{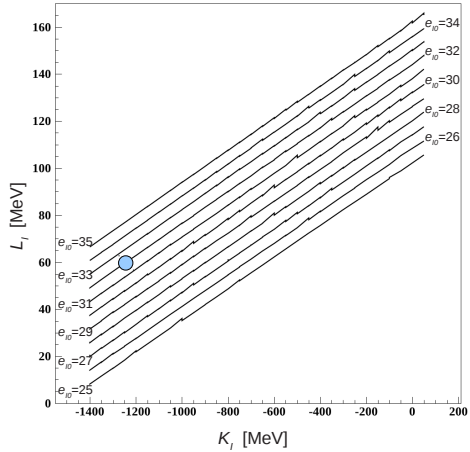}
  
\caption{Correlation between the slope, $L_{I}$, and curvature, $K_{I}$, of the
isospin symmetry energy for different Wigner energies, $e_{I0}$, obtained from
fitting the binding energies and rms radii of the $t$ and $^{3}$He. }

\label{fig2}
\end{figure}

Using a similar method as in the case of an alpha particle one can find
relationships between the parameters $e_{I0}$, $K_{I}$ and $L_{I}$, by requiring
a precise description of the binding energies and radii of the $t$ and $^{3}$He
particles. Here, the values of $e_{I0}$ and $K_{I}$ were varied independently on
a grid and the value of $L_{I}$ was found from the minimization of
(\ref{eq:ch2}). The results of such calculations are shown in Fig. \ref{fig2}.

The figure presents a relationship between the parameters $e_{I0}$, $L_{I}$ and
$K_{I}$. Selection of two of them allows to determine the third one. It should
be stressed that the above correlations have been obtained as a result of the
procedure reproducing the $t$ and $^{3}$He ground state properties with the
parameters $\rho_{0}=0.185$ fm$^{-3}$, $K_{0}=270$ MeV, $\lambda=0.4$ and
$e_{00}=-16.79$ MeV fixed from the ground state properties of an alpha particle
(Fig. \ref{fig1}). Three other parameters: $e_{ii0}$, $L_{ii}$ and $K_{ii}$ were found to
be almost independent of $e_{I0}$ and $K_{I}.$

\subsection{Discussion of the parameter values}

The final values of the EoS parameters adopted in this work have been obtained
by taking into account also the binding energy and size of a deuteron. The
values have been summarized in Tables \ref{tbl:eos1} and \ref{tbl:eos2}. For
this example set of values the model reproduces exactly the experimental values
of the rms radii and of the binding energies of the $d$, $t$, $^{3}$He and
$\alpha$ particles (see also Figs. \ref{fig4} and \ref{fig6}).

From the point of view of the present work, which focuses on the nuclear systems
with even and equal numbers of protons and neutrons (i.e. the spin-balanced
matter), the parameters: $e_{ii0}$, $L_{ii}$, $K_{ii}$ and $e_{ij0}$, $L_{ij}$,
$K_{ij}$ from the expansion (\ref{eq:eos_4-1}) are irrelevant. However, they do
play an important role when constructing the ground state configurations and
determining the properties of nuclei composed of asymmetric matter. 

As far as the interactions between neutrons and protons are concerned, the
Coulomb repulsion between protons causes that the proton matter tends to locate
itself in the outer regions of the nucleus. This leads to differences between
the neutron and proton density distributions, even for symmetric nuclei, and
triggers additional interactions associated with the isospin symmetry energy
(proton or neutron skins can be created). This effect can be negligible in case
of light nuclei, while for heavier ones it may be more significant. Therefore,
for heavier nuclei, one has to use a reliable set of isospin symmetry energy
parameters $e_{I0}$, $L_{I}$ and $K_{I}$, even for symmetric nuclear matter.

As has been shown above, the binding energies and the rms radii of $t$, $^{3}$He
and $\alpha$ impose certain constraints on the parameters of the EoS. In case of
the $e_{00}$, $\rho_{0}$, $K_{0}$ and $\lambda$ parameters, once two of them are
fixed by some other constraints or means, or from some other evidences, the
remaining ones can be determined using the obtained correlations (Fig.
\ref{fig1}). 
Here, the compressibility parameter of symmetric matter $K_{0}$ should fall
within the range of about 250$\pm$50 MeV according to the recent experimental
constraints (flow interpretation \cite{dani02}, monopole vibrations
\cite{pie04,col04}, subthreshold kaon production \cite{sturm01,hart06}). It has
been fixed to 270 MeV in the present approximation. The volume energy parameter
$e_{00}$ should take the value of about -16 MeV according to the LDM. In the
present approximation it has been assumed to have the value of -16.79 MeV. The
values of the remaining two parameters, $\rho_{0}$ and $\lambda$ have been then
uniquely determined from the correlation plot of Fig. \ref{fig1}.

In case of the symmetry energy, the choice of any two parameters out of
$e_{I0}$, $L_{I}$,  $K_{I}$, allows to determine the third one (Fig. \ref{fig2}). The
compilation \cite{Bali2013} of the latest results of the nuclear physics
experiments and astrophysical observations suggests that the value of the
symmetry energy constant, $e_{I0}$, and of the slope parameter, $L_{I}$, should
fall within a range of about 32 $\pm$ 1 MeV and of about 59 $\pm$ 17 MeV,
respectively. In the present approximation the values of these parameters have
been assumed to be 32 and 59.405 MeV, respectively. These values yield the value
of $K_{I}$ to be about -1250 MeV, according to the correlation plot of Fig. \ref{fig2}.
The experimental constraint of \cite{Centelles2009} on the symmetry energy
compressibility parameter, $K_{I}$, yields its value to be about $-50 \pm$ 200
MeV. Another compilation \cite{bao} locates the theoretical values of $K_{I}$
between $-400$ and +466 MeV and its experimental values in a range from $-566
\pm$ 1350 MeV to 34 $\pm$ 159 MeV. Thus, this parameter is least constrained
experimentally and theoretically so far. Its value obtained within the model,
preferring very soft symmetry energy, is consistent with the negative
experimental constraint, thanks to its large error bar. However, we should
stress that the adopted set of values is not at all unique. As seen from the
correlation plots, some other choice of parameters is equally well possible. It
should be also emphasized that the assumed value of $\rho_{0}$ equal to 0.185
fm$^{-3}$, which fits well properties of the LCPs, and is
higher than the commonly used values of $0.16-0.17$ fm$^{-3}$, may influence
values of the remaining parameters. 

In order to describe the exotic nuclei and provide more precise parameters of
the symmetry energy, definitely some more advanced parameter searches will be
needed, involving e.g. fitting the properties of the neutron skin nuclei and
also of those approaching the proton drip line across the chart of the nuclides,
but this is beyond the scope of the present work.

\begin{table}[h]
  \centering 
\begin{tabular}{|c|c|c|c|c|c|c|}
\hline 
$\rho_{0}$  & $e_{00}$  & $K_{0}$  & $e_{I0}$  & $L_{I}$  & $K_{I}$  & $\lambda$\tabularnewline
\hline 
\hline 
$\left[\frac{1}{fm^{3}}\right]$  & $\left[MeV\right]$  & $\left[MeV\right]$  & $\left[MeV\right]$  & $\left[MeV\right]$  & $\left[MeV\right]$  & $1$\tabularnewline
\hline 
0.185  & -16.79  & 270  & 32  & 59.405  & -1250  & 0.4\tabularnewline
\hline 
\end{tabular}\caption{Parameters of the standard EoS which together with a parameter of
the surface energy $\lambda$ are reproducing the ground state energy
and rms radius of the alpha particle, $t$ and $^{3}$He. \label{tbl:eos1}}
\end{table}
\begin{table}[h]
  \centering 
\begin{tabular}{|c|c|c|c|c|c|}
\hline 
$e_{ii0}$  & $L_{ii}$  & $K_{ii}$  & $e_{ij0}$  & $L_{ij}$  & $K_{ij}$ \tabularnewline
\hline 
\hline 
$\left[MeV\right]$  & $\left[\frac{1}{fm^{3}}\right]$  & $\left[MeV\right]$  & $\left[MeV\right]$  & $\left[MeV\right]$  & $\left[MeV\right]$ \tabularnewline
\hline 
82.37  & 242.48  & 19.909  & -1.5  & 182.  & 1210. \tabularnewline
\hline 
\end{tabular}\caption{Parameters of the spin interaction chosen in such 
a way that together
with parameters from Table 1 they reproduce binding energies and sizes
of $d,\: t$ and $^{3}$He. \label{tbl:eos2}}
\end{table}

A concise summary of the procedure of fixing the parameter values is presented
in the Appendix.

\section{Energy and size of nuclei with equal and even number of neutrons and
protons}

A method of finding the ground state configuration based on minimization of the
average value of the liquid drop-like Hamiltonian (\ref{eq:Ham_int}) with
respect to the wave function parameters $\left\langle
\mathbf{r}_{k}\right\rangle $ and $\sigma_{k}(r)$ with the $\left\langle
\mathbf{p}_{k}\right\rangle =0$ constraint, is described in
\cite{Sos_2014,Sos_2010}. The ground state configuration parameters determined
this way indicate that the corresponding wave function has some numbers of
centers, which can be interpreted as the centers of alpha-like clusters in the
nuclear matter. Some features of these ground state configurations have been
discussed in \cite{Sos_2014} without taking into account interactions and
correlations among clusters. In the present approach modification of the ground
state configurations due to the presence of the alpha-alpha interactions will be
investigated.

The main aim is to identify such symmetries of the ground state configuration,
and thus of the corresponding wave function, which could be responsible for the
experimentally observed changes of the binding energies and rms radii of the
$n$-alpha nuclei as a function of $n$. We would like to check whether it is
possible to identify specific configurations related to the magic numbers. In
other words, we would like to verify whether the shell corrections necessary to
describe deviations from the smooth predictions of the LDM around
the magic numbers, could be interpreted in terms of a change of the ground state
configuration resulting from the corrections to the Hamiltonian related to the
cluster-cluster interactions.  As will be justified in the following, the ground
state configurations corresponding to the local minima of the Hamiltonian can be
classified according to the nuclear charge, $Z$, in the following way:

\noindent \begin{tabular}{rr@{$Z$}lp{9cm}}
i.   &       &   $\leq8$  & 
The structures are identical with the structures proposed in \cite{Hafs1938}
and also in \cite{Sos_2014} and are characterized by no alpha cluster
in the center of the system.\\

ii.  & $ 8<$ &   $\leq20$ & 
A core represented by one alpha-like cluster appears in the
center of the system.\\

iii. & $20<$ &   $\leq28$ & 
A core with a mass corresponding to two alpha particles appears 
in the center.\\
\end{tabular}

%
%

\begin{figure}[h]
  \centering 
  \includegraphics[scale=0.5]{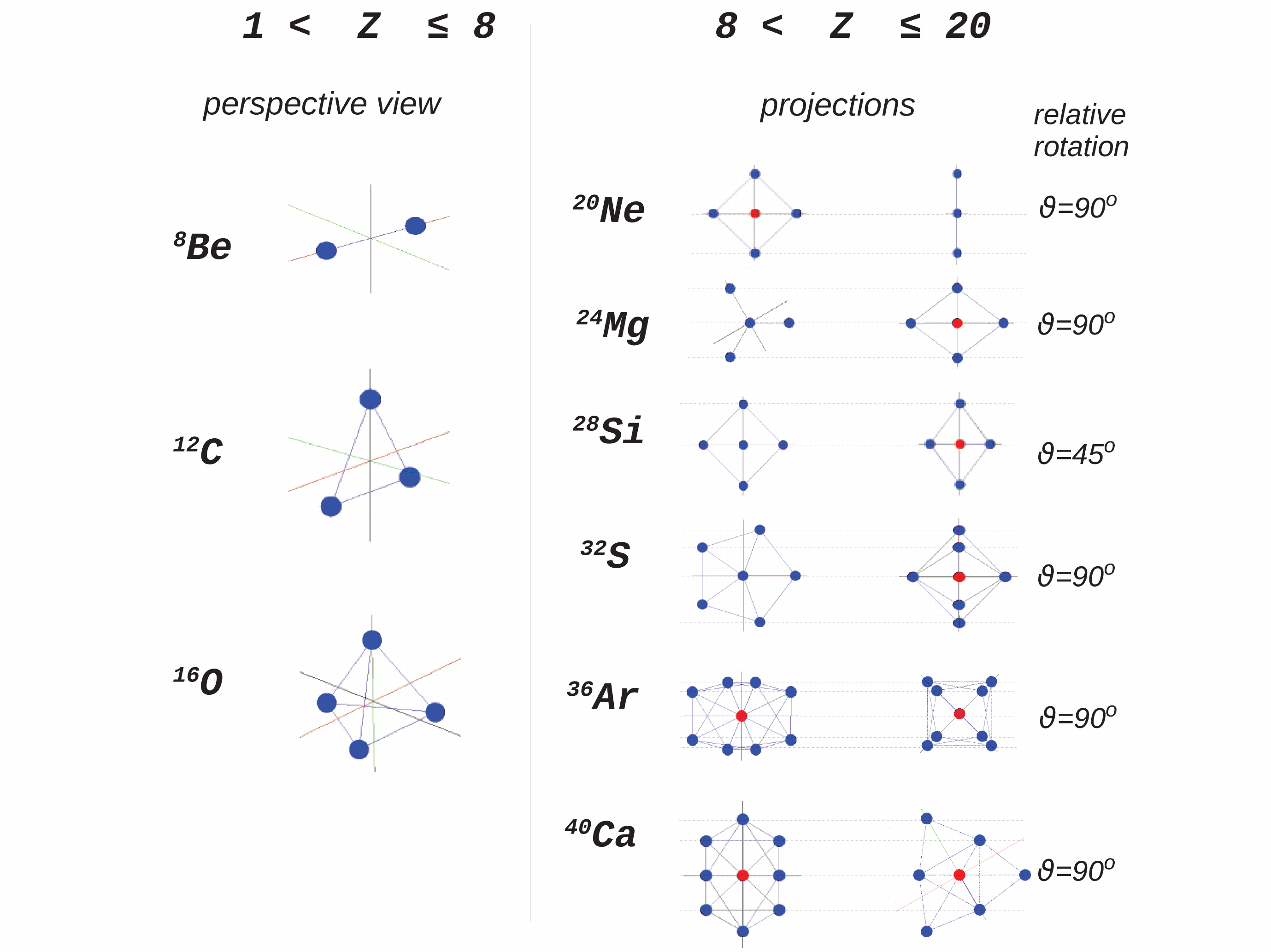} 
\caption{Proposed structures of $n$-alpha nuclei. These structures ensure high
level of symmetry for a given number of alpha particles, (see \cite{Web}).
The density distribution of the central cluster (the red one) has quite a big 
spread and therefore it cannot be identified as an alpha particle.}
\label{fig3}
\end{figure}

\noindent The structures for cases i and ii are visualized in Fig. \ref{fig3}
(see also \cite{Web}). 

The structures presented in Fig. \ref{fig3} result from the analysis of a large
number of configurations obtained within a procedure of minimizing the
Hamiltonian (\ref{eq:Ham_int}) when starting from a random configuration (see
$^{12}$C example in Fig. \ref{figmin}). Majority of thus obtained configurations
could be classified as those presented in the previous paper \cite{Sos_2014},
with alpha-like clusters located mostly at the surface and with a moderate level
of symmetry. Binding energies for these structures followed the smooth LDM trend
and could not reproduce the ``fine'' structure of binding energies around the
magic numbers. In some cases however, the minimization procedure did not yield
the global minimum, but ended up in some local one, just about 0.3 - 0.5 MeV
above the ground state energy of majority of configurations. Configurations
corresponding to these local minima were characterized by a higher level of
symmetry and possibly contained alpha-like cluster(s) in the middle.
Surprisingly, binding energy distribution for these structures did show
characteristic ``kinks'' around magic nuclei, despite some energy offset. These
specific structures are those presented in Fig. \ref{fig3}. 

The question that arose from the above observation was whether it might be
possible to convert the local minima into global ones, while keeping the highly
symmetric structures, by introducing the corrections to the Hamiltonian which
take into account the cluster-cluster interactions. The following sections try
to verify this hypothesis.

\subsection{Matter without alpha clusters}

The structures from Fig. \ref{fig3} have been used as structural constraints in
the minimization procedure by providing the starting points for the preformed
alpha-like clusters. In other words, the random initial configurations, as in
Fig. \ref{figmin}, have been replaced by those from Fig. \ref{fig3}. The actual
ground state configurations were determined through minimization of the liquid
drop-like Hamiltonian (\ref{eq:Ham_int}) by varying the mean positions and the
widths of the individual wave packets.

The following analysis is restricted to nuclei smaller or equal to $^{40}$Ca
which is the heaviest stable $n$-alpha nucleus. Beyond $Z=20$ the nuclei become
radioactive and there are no data about their sizes available.

\begin{figure}[h]
  \centering 
 \includegraphics[scale=0.4]{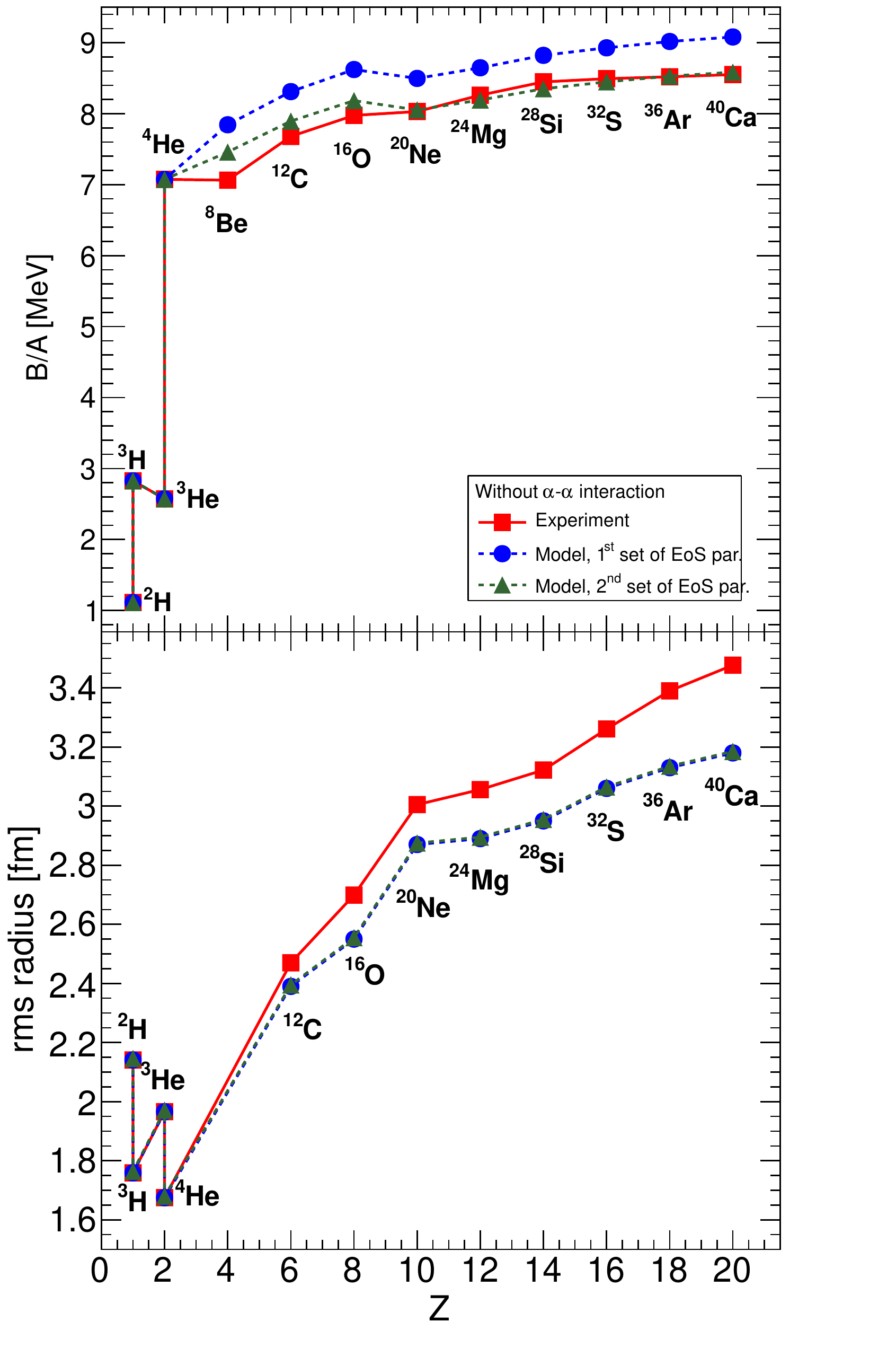}
\caption{The binding energies and rms radii of even-even nuclei ($N=Z$) 
obtained for the ground state wave functions and the Hamiltonian 
(\ref{eq:Ham_int}). The solid line connects the experimental points. Circles and
triangles represent model results for two sets of parameters of the EoS.}

\label{fig4}
\end{figure}

The liquid drop-like Hamiltonian (\ref{eq:Ham_int}) with the parameters from
Table 1 and Table 2 has been used together with structural constraints to obtain
the ground state configurations for each of the considered nuclei. The
minimization procedure provided the final parameters determining the ground
state wave function: $\left\langle\mathbf{r}_{i}\right\rangle$ and $\sigma_{i}$. Figure \ref{fig4} presents the obtained
ground state binding energies and rms radii as a function of the atomic number
$Z$. The experimental data and the model results are represented by the squares 
and the circles, respectively. As one can see the calculated binding energies
change their slope at the same place ($Z=8$) as the experimental ones and the
same concerns the radii. The overall calculated binding energies are somewhat
bigger by about 0.3 - 0.5 MeV/nucleon than the experimental ones and the
calculated radii are somewhat smaller, however both follow the experimental
trends. Properties of particles up to $^{4}$He are reproduced perfectly well,
because they served as the constraints in the parameter fitting procedure.

In order to check the sensitivity of the results on the volume energy parameter
$e_{00}$ similar calculations for $e_{00}$ equal to -16.1 MeV instead of -16.79
MeV have been performed. The results are presented with a dotted green line in
Fig. \ref{fig4}. In case of the rms radii there is almost no difference
observed. The binding energies still deviate from the experimental values, but
what is important, similar changes of the slopes in theoretical and experimental
trends are observed. Results of other trial calculations in which all the
remaining parameters from $e_{0}$ to $\lambda$ were varied, did not yield better
reproduction of the experimental values with the use of the liquid drop form of
the Hamiltonian together with the structural constraints i-iii. Better
reproduction could only be obtained with the corrections to the Hamiltonian
discussed in the next section. The structural constraints assure configurations
for which global minima of energy are obtained, provided that the alpha-alpha
interactions are taken into account. Otherwise, they correspond to some local
minima.

\subsection{Modification of the Hamiltonian due to alpha clusters.}

\begin{figure}[ht]
  \centering 
 \includegraphics[scale=0.32]{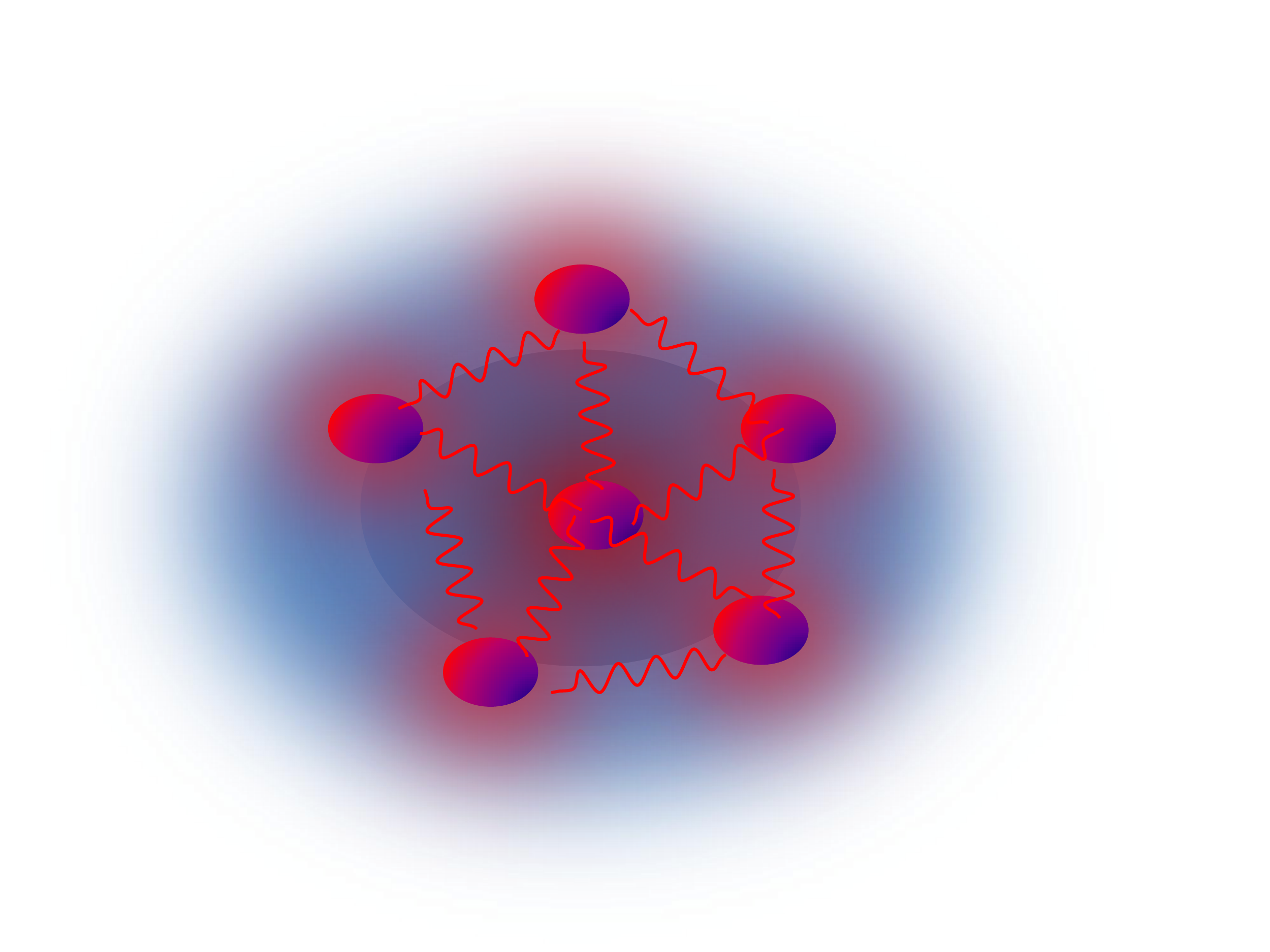}
\caption{Visualization of the additional interactions between alpha clusters
due to the many body interactions.}

\label{fig5}
\end{figure}

Let us consider now formation of the nucleus by randomly located nucleons
remaining within the range of their mutual interaction. Due to the Hamiltonian
(\ref{eq:Ham_int}) nucleons are forming groups in such a way that the energy of
the system is minimized. At certain density of the matter alpha-like clusters
can be formed as a result of the spin-isospin interactions of nucleon groups. As
alpha clusters are formed additional extra interactions connected with multi
body effects can appear having a strong influence on the final structure of the
system. Fig. \ref{fig5} illustrates a ground state of such a system. One can see that
inside a matter drop alpha clusters appear and interactions between them are
changing the energy of the system. Below, corrections to the Hamiltonian arising
from these extra interactions will be defined.

Assuming that alpha-like clusters appear in the structure of the nucleus, one
can expect two kinds of minor adjustments of the Hamiltonian, which should
describe the essential part of the $\alpha-\alpha$ interaction.

The first correction corresponds to the possible interaction caused by the
appearance of specific many-body forces. Such many-body effect can be
interpreted as a modification of the type and numbers of exchange bosons in the
interaction process. It appears possible that, in such an exchange the charge
currents are strongly limited because they change the energy of the considered
clusters. In this case, we can expect additional term in the interaction
potential which gives the repulsive forces between groups of nucleons forming a
cluster of alpha-like particles (see Fig. \ref{fig3}).

Here we test this hotfix in a form given by the product of the harmonic
oscillator potential $V_{\alpha\alpha}\left(d_{ij}\right)$ and probabilities
$P_{\alpha}\left(i\right),\: P_{\alpha}\left(j\right)$ that given
four nucleons (groups named $i$ or $j$) form the alpha clusters:
\begin{equation}
\left\langle \Phi\left|\Delta H_{1}\right|\Phi\right\rangle =\sum_{i\neq j}P_{\alpha}\left(i\right)P_{\alpha}\left(j\right)V_{\alpha\alpha}\left(d_{ij}\right)\label{eq:DH1}
\end{equation}
 where potential $V_{\alpha\alpha}\left(d_{ij}\right)$ is taken in
the form:

\begin{equation}
V_{\alpha\alpha}\left(d_{ij}\right)=\kappa\left(d_{ij}-d_{0}\right)^{2}\label{eq:vaa}
\end{equation}
 with $d_{ij}$ being the average distance between clusters $i$ and $j$,
and $P_{\alpha}\left(i\right)$ describing the probability that a group
of four nucleons $4n(i)$ forms an alpha cluster $i$. We assume that
$P_{\alpha}\left(i\right)$ can be expressed as:

\begin{equation}
P_{\alpha}\left(i\right)=\mathrm{exp}\left(-\nu\frac{\sigma_{4n(i)}-\sigma_{\alpha0}}{\sigma_{\alpha0}}\right)^{2}\label{eq:prob_DH1}
\end{equation}
 where $\sigma_{4n(i)}$ is the variance of the distribution describing
the probability of finding a nucleon in a group $4n\left(i\right)$
and $\sigma_{\alpha0}$ describes such a variance (experimentally
measured) for alpha particles. In the above formulas $\kappa,\: d_{0}$
and $\nu$ are free parameters. It is assumed that the potential is
equal to 0 when $d_{ij}>4$ fm with a shape close to the Van der Waals
one.

The second correction is related to the additional symmetry of the wave
function, which can be associated with the bosonic character of an alpha
cluster.

In the present approach nucleons and their clusters are distinguishable objects
(\ref{eq:wf}). For nucleons their energy connected with the fermionic motion is
taken into account by applying the energy density functional. For alpha
particles a relevant correction to energy should also appear and it is an
empirical one.

When calculating the minimum of the Hamiltonian by varying the parameters of the
wave function, the position of the alpha like cluster and its size and spread
are established.

These quantities can be different for different clusters depending on the
adopted structural constraint for a given nucleus. The exchange of clusters must
cause an increase of the energy of the system as a minimum of the Hamiltonian
defines definite spreads of clusters for their different positions.

For the indistinguishable alpha particles due to the symmetrization with respect
to the exchange of cluster positions there is no chance to relate cluster sizes
to their positions.

The second correction is an empirical one: 
\begin{equation}
\left\langle \Phi\left|\Delta H_{2}\right|\Phi\right\rangle =\mu N_{\alpha}\sum_{i=1}^{i=A}\left(\frac{\sigma_{i}(r)-\sigma_{0}(r)}{\sigma_{0}(r)}\right)^{2}\label{eq:DH2}
\end{equation}
 where $N_{\alpha}$ is the number of alpha clusters and $\mu$ is a
parameter. $\sigma_{i}(r)$ is the variance of the nucleon radius
and $\sigma_{0}(r)$ is the average variance of nucleons forming an alpha
particle.

Thus, this correction to the Hamiltonian describes an increase of the energy
caused by different sizes of clusters and it should vanish when the wave
functions for all the clusters correspond to alpha particles in their ground
states.

It seems that this correction could be related to the bosonic character of 
alpha particles. Both corrections are positive and in this way the binding
energy decreases. Assuming that parameters of the main part of Hamiltonian
(\ref{eq:Ham}) are established by properties of light nuclei $d$, $t$, $^{3}$He
and $\alpha$, the idea now is to define additional four parameters in order to
reproduce binding energies and sizes of nuclei starting from $^{8}$Be till
$^{40}$Ca by imposing the structural constraints.

The adopted values of these parameters are summarized in Table 3.

\begin{table}[h]
  \centering 
\begin{tabular}{|c|c|c|c|}
\hline 
$\mu$  & $\nu$  & $d_{0}$  & $\kappa$ \tabularnewline
\hline 
\hline 
$\left[MeV\right]$  & $\left[1\right]$  & $fm$  & $\left[MeV/fm^{2}\right]$ \tabularnewline
\hline 
1.  & 9.  & 3.77  & 5.2 \tabularnewline
\hline 
\end{tabular}\caption{Estimated parameters of corrections (\ref{eq:DH1}), and (\ref{eq:DH2})
to the Hamiltonian (\ref{eq:Ham_int}) coming from interactions between
alpha clusters. \label{tbl:a-a int}}
\end{table}

\begin{figure}[h]
  \centering 
\includegraphics[scale=0.4]{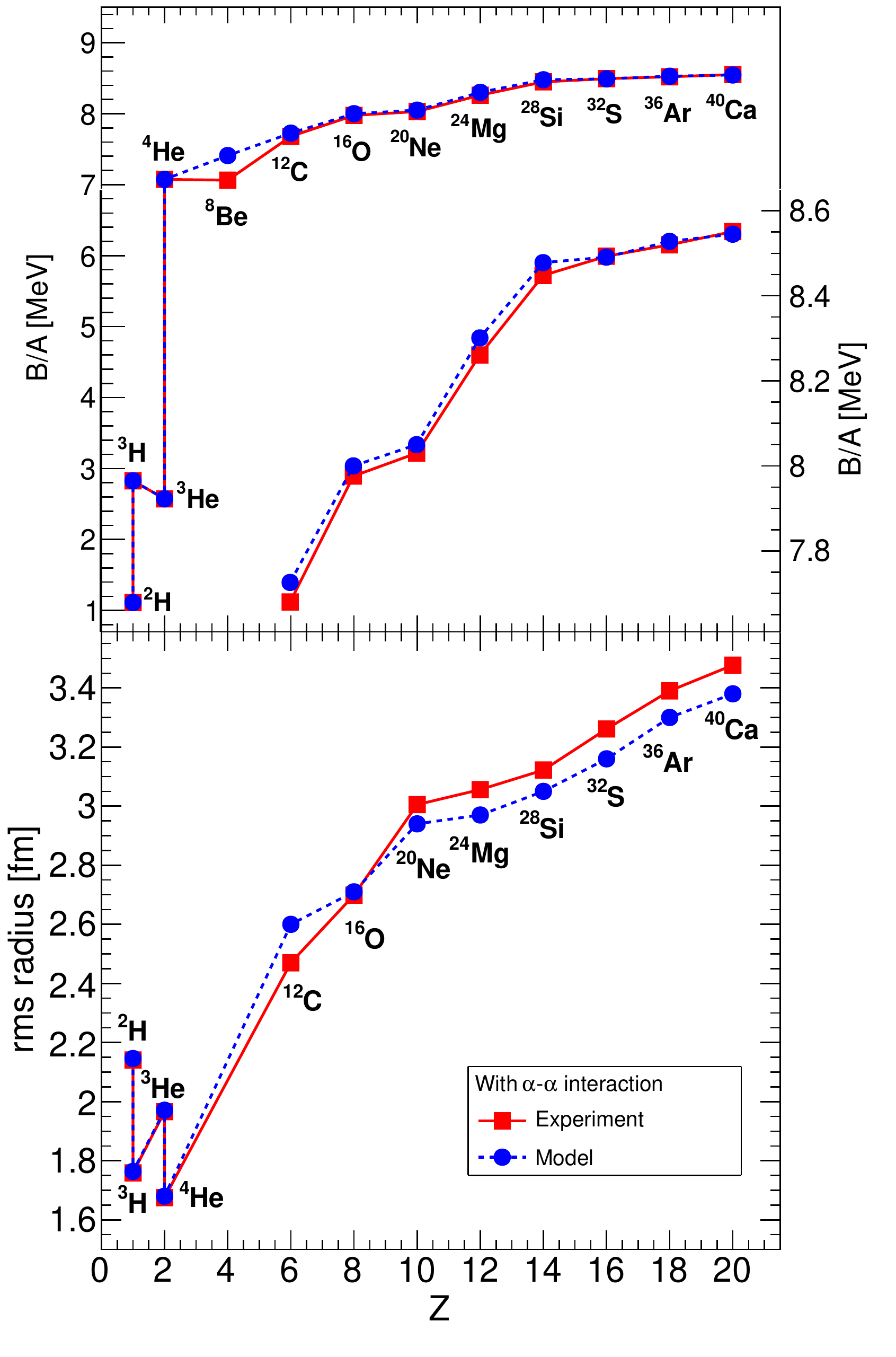}
\caption{Binding energies per nucleon and rms radii for the Hamiltonian
with the corrections (\ref{eq:DH1}) and (\ref{eq:DH2}). The right scale of
binding energies per nucleon in the
top panel refers to the magnified view for nuclei from $^{12}$C to $^{40}$Ca.
The solid (dashed) line connects the experimental (model) points.}

\label{fig6}
\end{figure}

The results for the full Hamiltonian, including the corrections due to the
alpha-alpha interactions are presented in Fig. \ref{fig6}. The upper part of the figure
shows the binding energies as a function of the atomic number $Z$. The bottom
part presents the rms radii. 

Much better agreement between the model results and the experiment can be
observed now, as compared to  Fig. \ref{fig4}. Some discrepancies can be still noticed
especially for $^{8}$Be for which the distance between the alpha particles is
pretty small and much smaller than $d_{0}$ in the harmonic potential
description. Again, the properties of particles up to $^{4}$He are reproduced
perfectly well, because they served as the constraints to fix the EoS parameters.

\begin{figure}
  \centering 
\includegraphics[scale=0.3]{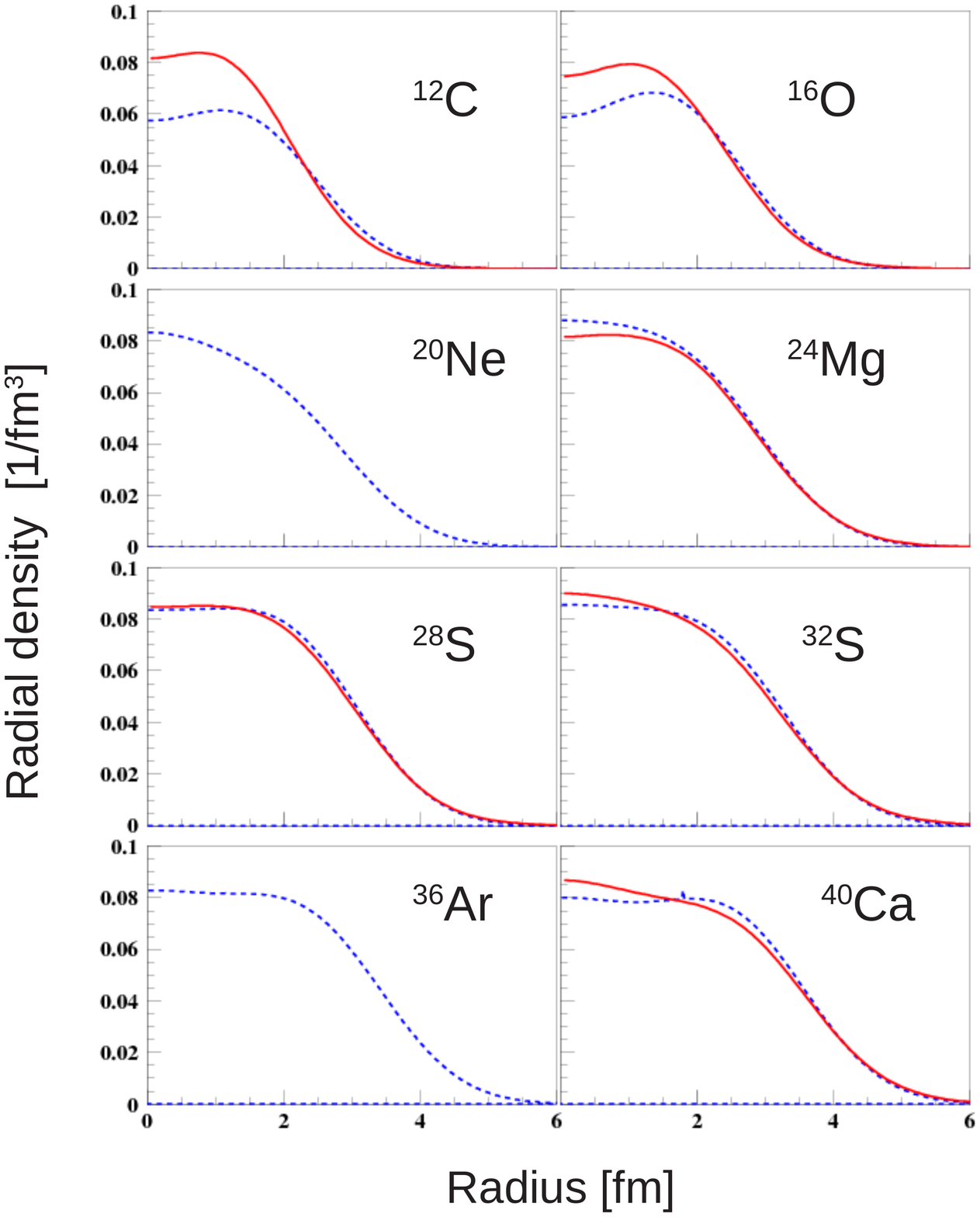}
\caption{Experimental (solid, red) and model (dashed, blue) density profiles. }

\label{fig7}
\end{figure}

Figure \ref{fig7} presents a comparison of the model and experimental density profiles
for the selected nuclei. The full (red) lines represent the experimental
profiles and the broken (blue) lines result from the model calculations.
Experimental density profiles were taken from \cite{Vries87}, however for
$^{20}$Ne and $^{36}$Ar reliable experimental data were not available.

\begin{figure}
  \centering 
\includegraphics[scale=0.3]{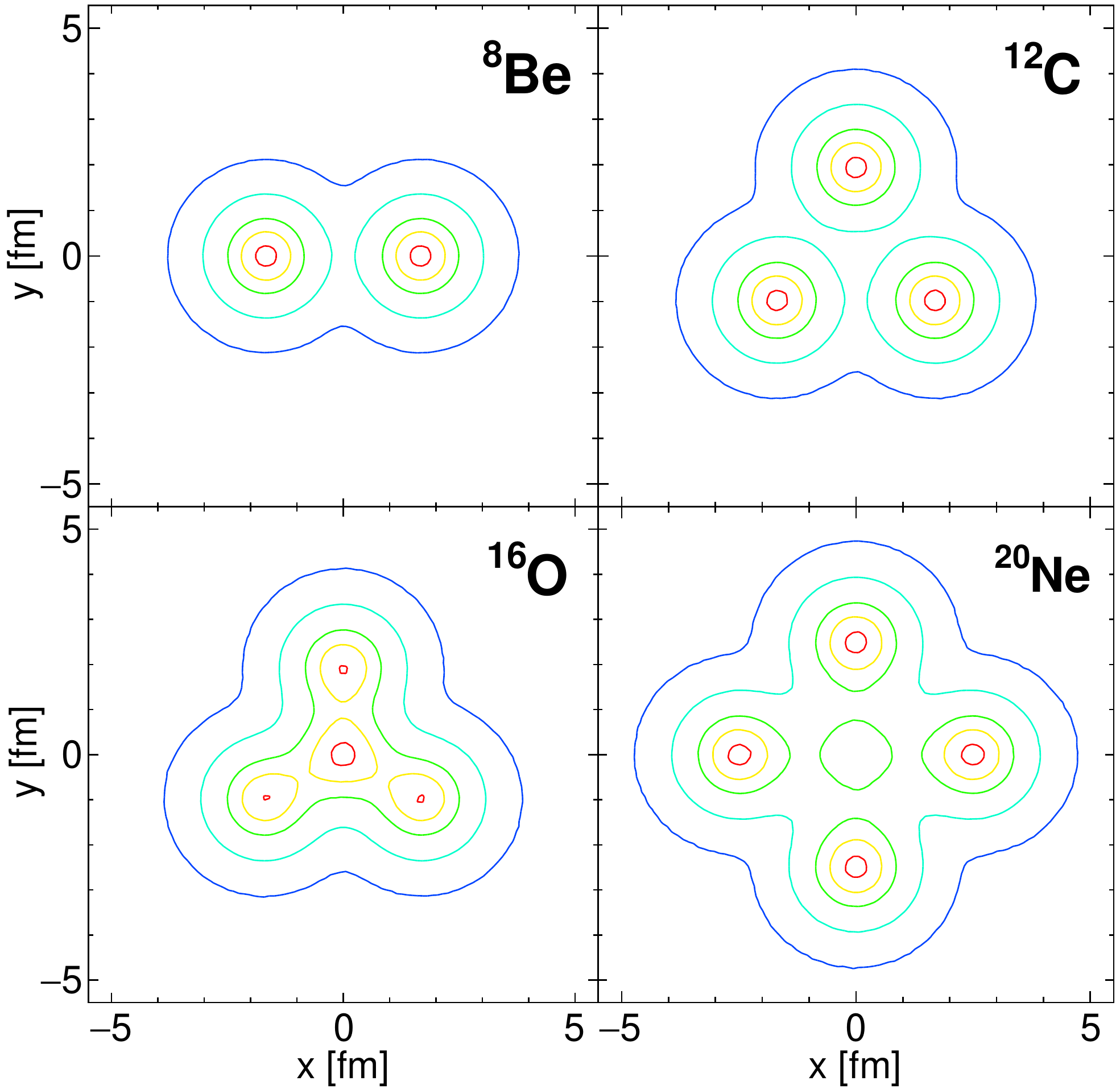}
\caption{Shapes of the model nuclei. The isodensity lines correspond to 1, 15, 
50, 75 and 95\% of the maximum density.}

\label{fig8}
\end{figure}

Figure \ref{fig8} shows the isodensity contour plots for the model ground state
configurations of $^{8}$Be, $^{12}$C, $^{16}$O and $^{20}$Ne. Please note that
apart from the $^{16}$O, which has a tetrahedral shape, all the other nuclei
have a planar geometry (cf. Fig. \ref{fig3}). Similar shape for $^{12}$C has been
obtained within the FMD \cite{feld97} and AMD \cite{kanada01} simulations. In
particular, the $^{20}$Ne case shows that the central $\alpha$-like cluster is
more diffused than the surface ones. More quantitatively, the standard
deviations of the density distributions of surface alphas are of the order of
1.02 fm, while for the core alpha they are of the order of  1.28 fm.

\begin{figure}[ht]
  \centering 
 \includegraphics[scale=0.4]{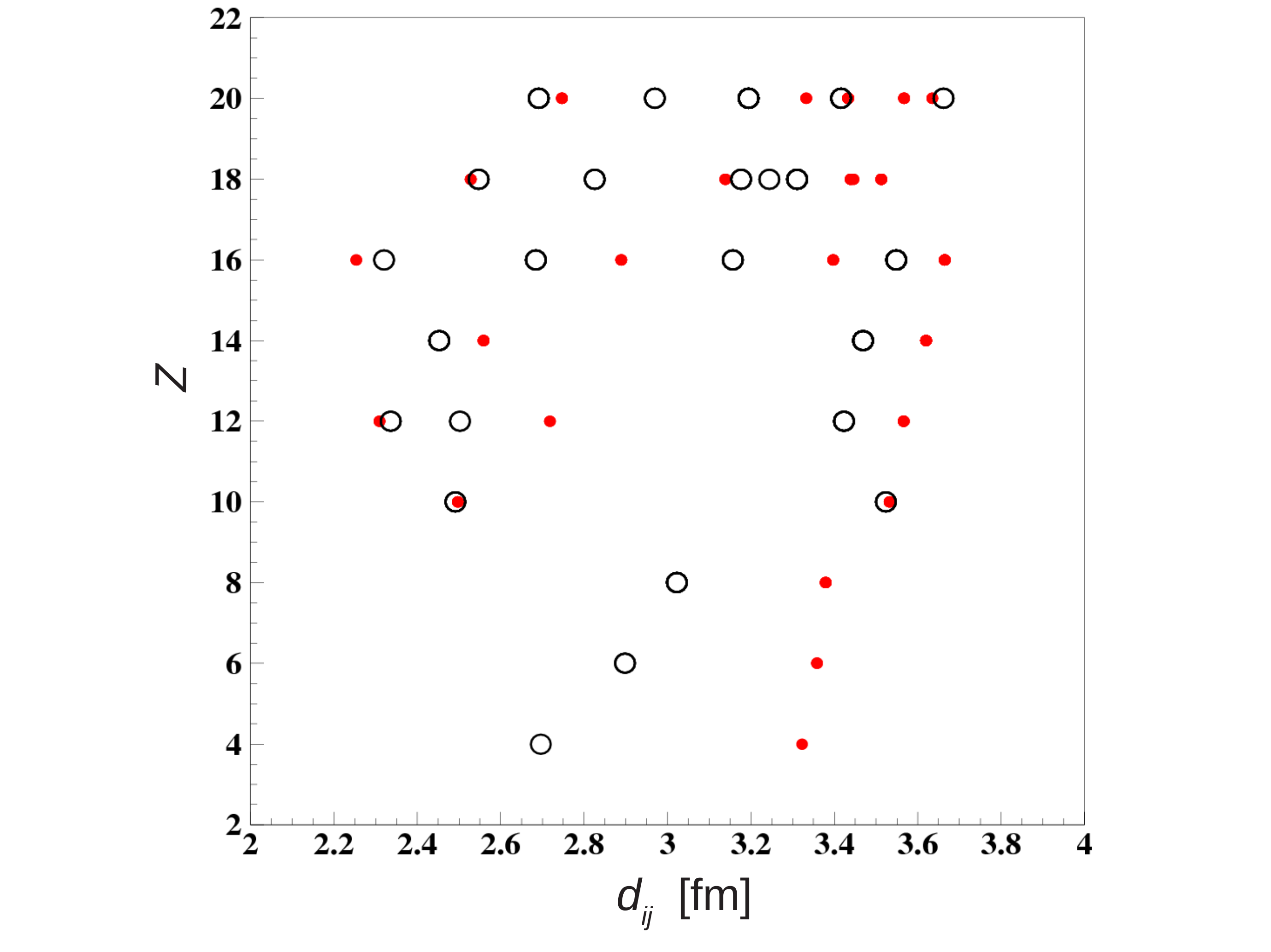}
 
\caption{The distribution of the distances between alpha clusters as a function
of Z. The open (closed) circles represent distances without (with) the
correction for alpha-alpha interaction in the Hamiltonian.}

\label{fig9}
\end{figure}

Figure \ref{fig9} presents a distribution of distances between alpha clusters for the
nuclei in question.

The distances between clusters from the ground state configurations obtained
with the liquid drop-like Hamiltonian without corrections are represented by
open circles. The distances obtained from the full Hamiltonian calculations are
represented by closed ones. As can be seen, in the latter case the distances are
grouping around 3.5 fm. This can denote that these clusters behave more like
rigid spheres.

\section{Summary and Conclusions}

A simple model of nuclei formed out of alpha-like clusters in which interactions
lead to strong correlations between nucleons has been presented. The model
interactions have been defined using a cubic approximation of the nuclear EoS.
The proposed EoS contains additional terms which take into account the spin and
isospin polarizations. These additional terms cause that nucleonic clusters,
with an enlarged binding energy, appear in the ground state configurations.
These clusters have full spin-isospin symmetry similarly as in \cite{Ono_92} and
in \cite{Feld_90}. For $n$-alpha nuclei these clusters can be identified as
alpha like structures. When these clusters are formed in regions of smaller
density, i.e. at the nuclear surface, then their sizes measured as the variance
of the density distribution become comparable to the alpha particle size. On
the other hand formation of the clusters in the central part of nuclei leads to
much bigger spreads of their density distributions and they can hardly be
identified as alpha particles.

It has been shown that by taking into account the liquid drop-like form of the
Hamiltonian in the minimization process it was not possible to describe the
subtle changes of the binding energies of nuclei. Therefore the liquid drop part
of the Hamiltonian has been supplemented by the corrections taking into account
the cluster-cluster interactions. The ground state configurations obtained from
the minimization of the full Hamiltonian with the imposed structural constraints
were found to reproduce very well the experimental binding energies and sizes of
the $n$-alpha nuclei and the available density profiles. In particular, the
model with the cluster corrections has been able to reproduce the "fine
structure"  of the ground state energy distribution (inset of Fig. \ref{fig6}).
In this respect it seems to be able to mimic the shell effects without explicit
spin-orbit term.

The interactions have been defined using a 12 parameter EoS approximation. The
thirteenth parameter has been added to control the variance of the field for
finite systems. The values of these parameters and correlations among them have
been constrained using the experimental binding energies and sizes of $d$, $t$,
$^{3}$He and $\alpha$ particles. Additional 4 parameters have been introduced in
the corrections to the liquid drop part of the Hamiltonian in order to account
for the interactions between the alpha-like clusters and to substantially improve
the model predictions. The differences between the calculated and measured
ground state energies for nuclei with Z$>$8 are about an order of magnitude
smaller than those obtained using the liquid drop model. Besides, the obtained
parameters describe the properties of the $t$, $^{3}$He and $\alpha$ very
accurately (practically error free).

It has also been demonstrated that correlations between nucleons in the phase
space can play an important role in the description of the ground states,
particularly of the ground states of nuclei with the same and even number of
neutrons and protons.

\section*{Acknowledgments}

Work supported by Polish National Science Centre (NCN), contract Nos.
UMO-2013/10/M/ST2/00624, UMO-2013/09/B/ST2/04064, IN2P3 08-128 and also by the Foundation for Polish Science - MPD program,
co-financed by the European Union within the European Regional Development
Fund.

Stimulating discussions with Professors Lucjan Jarczyk, Bogus\l{}aw Kamys,
Andrzej Magiera, Zbigniew Rudy, Roman P\l{}aneta, Andrzej Wieloch and
Janusz Brzychczyk are gratefully acknowledged. 

Special thanks to the anonymous referee for many suggestions that improved
the original version of the manuscript.

\section*{Appendix: Extraction of the EoS parameters}

The Hamiltonian (16) depends on 13 parameters and since the binding energies and
rms radii of the ground states of $d$, $t$, $^{3}$He and $\alpha$ provide only 8
constraints, 5 of the parameters have to be fixed by some other means.

The first step of the procedure to fix the EoS parameters consisted in
constraining the values of $K_{0}$, $\rho_{0}$, $e_{00}$ and $\lambda$ by the
binding energy and rms radius of an alpha particle. This step has been
summarized in the flow chart of Fig. \ref{fig10}.

\tikzstyle{startstop} = [rectangle, rounded corners, minimum width=3cm, minimum height=1cm,text centered, text width=8cm, draw=black]
\tikzstyle{io} = [trapezium, trapezium left angle=70, trapezium right angle=110, minimum width=1cm, minimum height=1cm, text centered, text width=3cm, draw=black]
\tikzstyle{io1} = [trapezium, trapezium left angle=70, trapezium right angle=110, minimum width=1cm, minimum height=1cm, text centered, text width=3cm, draw=black]
\tikzstyle{process} = [rectangle, minimum width=3cm, minimum height=1cm, text centered, text width=8cm, draw=black]
\tikzstyle{process1} = [rectangle, minimum width=3cm, minimum height=1cm, text centered, text width=3cm, draw=black]
\tikzstyle{decision} = [diamond, minimum width=3cm, minimum height=1cm, text centered, draw=black]
\tikzstyle{arrow} = [thick,->,>=stealth]

\begin{figure}[h!]
\centering
\begin{tikzpicture}[node distance=2cm]
\node (start) [startstop] {Select starting values of $\rho_{0}$ and $K_{0}$};
\node (in1) [io, below of=start, yshift=0.5cm] {Set starting values of $e_{00}$ and $\lambda$};
\node (pro1) [process, below of=in1, yshift=0.5cm] {Find ground state configuration 
(centroids and widths of Gaussian wave packets minimizing (16))};
\node (pro1b) [process, below of=pro1, yshift=0.5cm] {Calculate binding energy and rms radius of the ground state configuration and calculate $\chi^{2}$ of (17)};
\node (dec1) [decision, below of=pro1b, yshift=-0.5cm] {$\chi^{2}<10^{-7}$ ?};
\node (pro2a) [process, below of=dec1, yshift=-0.5cm] {Increment $\rho_{0}$ and/or $K_{0}$
while within predefined grid};
\node (pro2b) [process1, right of=dec1, xshift=2.7cm] {Vary $e_{00}$ and $\lambda$ };
\node (out1) [io1, below of=pro2a, yshift=0.5cm] {Produce Fig. \ref{fig1}};
\node (stop) [startstop, below of=out1, yshift=-0.0cm] {Assume values of 
$K_{0}$ and $e_{00}$ consistent with other experimental constraints and obtain the corresponding values of $\rho_{0}$ and $\lambda$
    from the correlation plot (Fig. \ref{fig1})};

\draw [arrow] (start) -- (in1);
\draw [arrow] (in1) -- (pro1);
\draw [arrow] (pro1) -- (pro1b);
\draw [arrow] (pro1b) -- (dec1);
\draw [arrow] (dec1) -- node[anchor=east] {yes} (pro2a);
\draw [arrow] (dec1) -- node[anchor=south] {no} (pro2b);
\draw [arrow] (pro2b) |- (pro1);
\draw [arrow] (pro2a.west) -- ++(-5mm,0) |- (in1);
\draw [arrow] (pro2a) -- (out1);
\draw [arrow] (out1) -- (stop);

\end{tikzpicture}

\caption{Flow chart of the procedure of fixing the $K_{0}$, $\rho_{0}$, $e_{00}$
and $\lambda$ parameters using the binding energy and the rms radius of an alpha
particle as a constraint. The details can be found in Sect. 3.\label{flow}}

\label{fig10}
\end{figure}
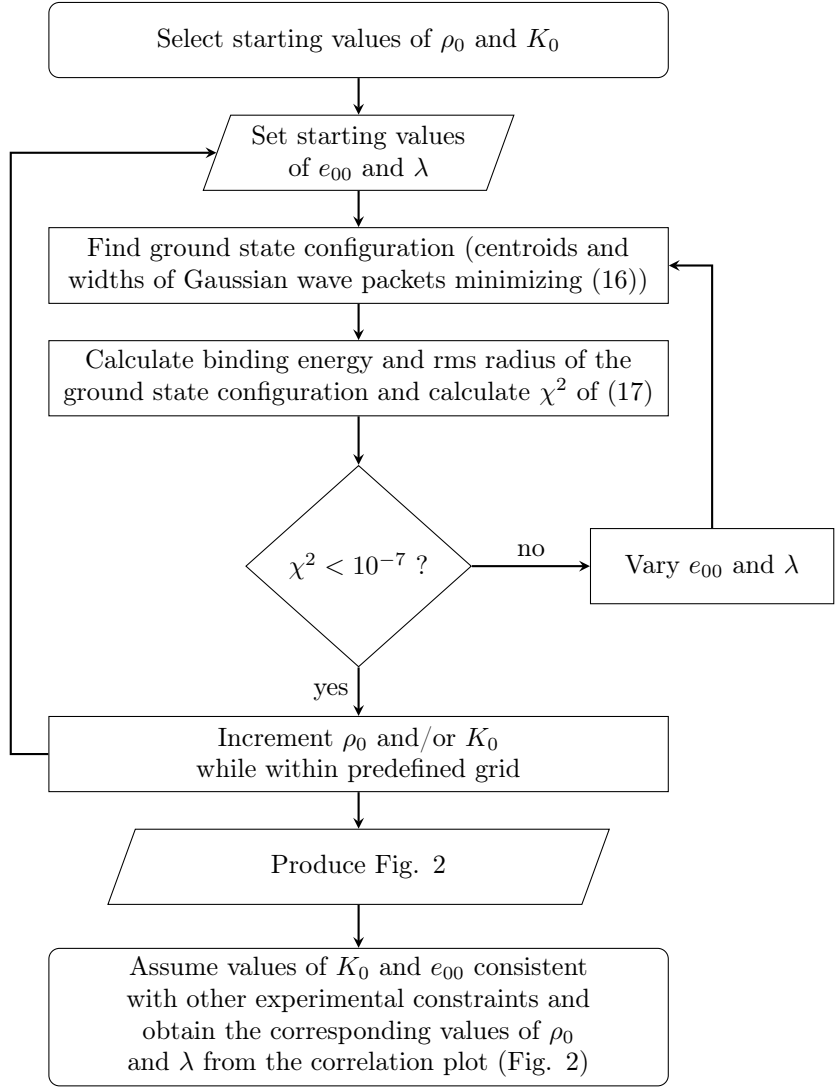

The compressibility parameter of symmetric matter $K_{0}$ and the volume energy
parameter $e_{00}$ have been fixed within the range of the recent experimental
constraints. The values of the remaining two parameters, $\rho_{0}$ and
$\lambda$ are then uniquely determined from the correlation plot of Fig. \ref{fig1}.
The remaining 9 parameters were irrelevant for the properties of an alpha
particle, which greatly simplified the corresponding EoS. 

In the second step, the parameters $e_{I0}$, $L_{I}$ and $K_{I}$ have been found
in an analogous way by using the binding energies and the rms radii of $t$ and
$^{3}$He as constraints in eq. (17), by fixing the previously determined
parameters $K_{0}$, $\rho_{0}$, $e_{00}$ and $\lambda$, and by disregarding the
remaining 6 parameters. 

Finally, the remaining 6 parameters $e_{ii0}$, $L_{ii}$, $K_{ii}$, $e_{ij0}$,
$L_{ij}$ and $K_{ij}$ have been found by minimizing (17) with the fixed 7
parameters determined in the previous two steps and using the binding energies
and the rms radii of $d$, $t$ and $^{3}$He as constraints. Fitting is done in an
analogous way as in the inner loop of the flow chart of Fig. \ref{fig10}, replacing
the $e_{00}$ and $\lambda$ parameters by the remaining 6 ones.

After fixing the 13 parameters of the liquid-drop-like Hamiltonian (16), the
additional four parameters of the $\alpha-\alpha$ interaction, $\mu$, $\nu$,
$d_{0}$ and $\kappa$, have been fixed by fitting the ground state binding
energies and rms radii of the nuclei from $^{8}$Be till $^{40}$Ca. Here, in
addition the structural constraints i. - iii. have been imposed in the fitting
routine, meaning that only the widths of the nucleon wave packets and the
inter-$\alpha$-like clusters distances could be varied, preserving the imposed
configuration.

The procedure of finding the ground state configurations which minimize the
model Hamiltonian is described in Sect. 4 of \cite{Sos_2010}, see also Fig.
\ref{figmin}.


\newpage

\begin{thebibliography}{References}
\bibitem{Bethe1936} H.A. Bethe, R.F. Bacher, Rev. Mod. Phys. \textbf{8},
82\textbf{ }(1936)

\bibitem{Hafs1938} L.R. Hafstad, E. Teller, Phys. Rev. \textbf{54},
681 (1938)

\bibitem{Web} http://nz21-33.ifj.edu.pl/clusters/

\bibitem{Freer2010} M. Freer (2010), Scholarpedia, 5(6):9652

\bibitem{vonoertzen06} W. von Oertzen, Eur. Phys. J. A \textbf{29}, 133 (2006)

\bibitem{Sos_2014} Z. Sosin, J. Kallunkathariyil, Acta Phys. Polon.
\textbf{B45}, 925\textbf{ }(2014)

\bibitem{Feld_90}H. Feldmeier, Nucl. Phys. \textbf{A515}, 147 (1990)

\bibitem{Ono_92} A. Ono \emph{et al.}, Prog. Theor. Phys. \textbf{87},
1185 (1992)

\bibitem{Kanada12} Y. Kanada-En'yo \emph{et al.}, Prog. Theor. Phys. \textbf{01A202},
(2012)

\bibitem{wile1}
L.~Wilets, \emph{et al.},
{\em Nuc. Phys. \/ {\bf A282}},~341(1977) 

\bibitem{dors1} 
C.~Dorso, S.~Duarte and J.~Randrup, {\em Phys. Lett. \/ {\bf 188B}},~287(1987)

\bibitem{hori1} 
H.~Horiuchi, A.~Ohnishi and T.~Maruyama, Proc. of the $6^{\mbox{th}}$
International Conference on Nuclear Reaction Mechanisms, Varenna,
June 10-15, 1991, ed. E.~Gadioli


\bibitem{boal1} 
D.H.~Boal and J.N.~Glosli, Phys. Rev. C \textbf{38},~1870 (1988) 

\bibitem{maru1} T.~Maruyama \emph{et al.}, {\em Phys. Rev. C\/ {\bf 53}},~297(1996)

\bibitem{fink} J. Fink \emph{et al.}, Z. Phys. A, 323, 189 (1986)


\bibitem{Aich91} J. Aichelin, Phys. Rep. \textbf{202,} 233 (1991)


\bibitem{Dirac81} P. Dirac, The Principles of Quantum Mechanics.
Oxford science publications (Oxford University Press, 1981)

\bibitem{Sos_2010} Z. Sosin, Int. J. Mod. Phys. E. \textbf{19}, 759 (2010)

\bibitem{bao} Bao-An Li \emph{et al.}, Int. J. Mod. Phys. E. \textbf{07}, 147 (1998)




\bibitem{dani02}P. Danielewicz \emph{et al.}, Science \textbf{298}, 1592 (2002)

\bibitem{pie04}J. Piekarewicz, Phys. Rev. C \textbf{69}, 041301 (2004)

\bibitem{col04}G. Col\`{o}  \emph{et al.}, Phys. Rev. C \textbf{70}, 024307 (2004)

\bibitem{sturm01}C. Sturm \emph{et al.}, Phys. Rev. Lett. \textbf{86}, 39 (2001)

\bibitem{hart06}Ch. Hartnack \emph{et al.}, Phys. Rev. Lett. \textbf{96}, 012302 (2006)

\bibitem{Bali2013}Bao-An Li and Xiao Han, Phys. Lett. B \textbf{727},
276 (2013)

\bibitem{Centelles2009}M. Centelles \emph{et al.}, Phys. Rev. Lett. \textbf{102},
122502 (2009)

\bibitem{Vries87} H. De Vries, C. W. De Jager, and C. De Vries
Atomic Data and Nuclear Data Tables \textbf{36}, 495 (1987) 

\bibitem{feld97} H. Feldmeier and J. Schnack, Prog. Part. Nucl. Phys. 
\textbf{39}, 393 (1997)

\bibitem{kanada01} Y. Kanada-En'yo and H. Horiuchi, Prog. Theor. Phys. Suppl.
No. \textbf{142}, 205 (2001)

\end{thebibliography}
\end{document}